**Human-Centered Artificial Intelligence: Reliable, Safe & Trustworthy**

Ben Shneiderman, ben@cs.umd.edu, Draft February 23, 2020, version 29
University of Maryland, College Park, MD USA

**Abstract**

Well-designed technologies that offer high levels of human control and high levels of computer automation can increase human performance, leading to wider adoption. The Human-Centered Artificial Intelligence (HCAI) framework clarifies how to (1) design for high levels of human control and high levels of computer automation so as to increase human performance, (2) understand the situations in which full human control or full computer control are necessary, and (3) avoid the dangers of excessive human control or excessive computer control. The methods of HCAI are more likely to produce designs that are Reliable, Safe & Trustworthy (RST). Achieving these goals will dramatically increase human performance, while supporting human self-efficacy, mastery, creativity, and responsibility.

1. **Introduction**

This paper opens up new possibilities by way of a two-dimensional framework of Human-Centered Artificial Intelligence (HCAI) that separates levels of automation/autonomy from levels of human control. The new goal is to seek **high levels of human control AND high levels of automation**, which is more likely to produce computer applications that are Reliable, Safe & Trustworthy (RST). Achieving this goal, especially for complex poorly understood problems, will dramatically increase human performance, while supporting human self-efficacy, mastery, creativity, and responsibility.

The traditional belief in computer autonomy is compelling for many artificial intelligence (AI) researchers, developers, journalists, and promoters. The goal of computer autonomy was central in Sheridan and Verplank's (1978) ten levels from human control to computer automation/autonomy (Table 1). Their widely cited one-dimensional list continues to guide much of the research and development, suggesting that increases in automation must come at the cost of lowering human control. Shifting to HCAI could liberate design thinking so as to produce computer applications that increase automation, while amplifying, augmenting, enhancing, and empowering people to innovatively apply systems and creatively refine them.

| Level | Description |
|---|---|
| High | 10. The computer decides everything and acts autonomously, ignoring the human. |
|  | 9. The computer informs the human only if it, the computer, decides to. |
|  | 8. The computer informs the human only if asked, or |
|  | 7. The computer executes automatically, then necessarily informs the human, and |
|  | 6. The computer allows the human a restricted time to veto before automatic execution, or |
|  | 5. The computer executes that suggestion if the human approves, or |
|  | 4. The computer suggests one alternative, or |
|  | 3. The computer narrows the selection down to a few, or |
|  | 2. The computer offers a complete set of decision/action alternatives, or |
| Low | 1. The computer offers no assistance; the human must take all decisions and actions. |

Table 1: Summary of the widely cited, but mind-limiting 1-dimensional Sheridan-Verplank levels of automation/autonomy (Parasuraman et al., 2000)

Sheridan & Verplank's ten levels of automation/autonomy have been widely influential, but critics suggested refinements such as the four *stages* of automation: (1) information acquisition, (2) analysis of information, (3) decision or choice of action, and (4) execution of action (Parasuraman, Sheridan & Wickens, 2000). These stages refine discussions of each of the levels, but the underlying message is that the goal is full automation/autonomy.

Even Sheridan (2000) commented with concern that "surprisingly, the level descriptions as published have been taken more seriously than were expected" (see Hoffman & Johnson (2019) for a detailed history). However, in spite of the many critiques, the 1-dimensional levels of automation/autonomy, which only represents situations where increased automation must come with less human control, is still widely influential. For example, the US Society of Automotive Engineers adopted the unnecessary trade-off in its six levels of autonomy for self-driving cars (SAE, 2014; Brooks, 2017) (Table 2).

| Level | Description |
|---|---|
| 5. | **Full autonomy**: equal to that of a human driver, in every driving scenario. |
| 4. | **High automation:** Fully autonomous vehicles perform all safety-critical driving functions in certain areas and under defined weather conditions. |
| 3. | **Conditional automation:** Driver shifts "safety critical functions" to the vehicle under certain traffic or environmental conditions. |
| 2. | **Partial automation:** At least one driver assistance system is automated. Driver is disengaged from physically operating the vehicle (hands off the steering wheel AND foot off the pedal at the same time). |
| 1. | **Driver assistance:** Most functions are still controlled by the driver, but a specific function (like steering or accelerating) can be done automatically by the car. |
| 0. | **No Automation:** Human driver controls all: steering, brakes, throttle, power. |

Table 2: Persistent, but still misleading, 1-dimensional thinking about levels of autonomy for self-driving cars (SAE, 2014; Brooks, 2017)

Critics of autonomy have repeatedly discussed the ironies (Bainbridge, 1983), deadly myths (Bradshaw et al., 2013; Mindell, 2015), conundrums (Endsley, 2017), or paradoxes (Hancock, 2017) of autonomy. A common point is that humans have to spend more effort monitoring autonomous computers because they are unsure of what it will do, often leading to inferior performance (Blackhurst et al., 2011; Strauch, 2017).

Bradshaw et al. (2013) caution designers who "have succumbed to myths of autonomy that are not only damaging in their own right but are also damaging by their continued propagation ... because they engender a host of other serious misconceptions and consequences." The U.S. Defense Science Board (Murphy and Shields, 2012) report describes the ten levels of automation/autonomy this way: "though attractive, the conceptualization of levels of autonomy as a scientific grounding for a developmental roadmap has been unproductive." A later report described many opportunities and dangers (Defense Science Board 2016), while cognitive science researchers (Hoffman et al., 2016) point to the failures and costs of autonomous weapons.

The problems brought by autonomy are captured in Robin Murphy's Law of autonomous robots: "any deployment of robotic systems will fall short of the target level of autonomy, creating or exacerbating a shortfall in mechanisms for coordination with human problem holders" (Woods et al., 2004). High-profile disasters from excessive autonomy in new technologies include the Patriot missile system shooting down two friendly aircraft during the Iraq War (Blackhurst et al., 2004) and the Boeing 737 MAX crashes (Nicas et al., 2019).

Fortunately, there is growing awareness by leading artificial intelligence researchers and developers that human-centered designs are needed (Jordan, 2018; Li, 2018). Equally important is the encouragement of creativity researchers to integrate creativity support features in technology (Candy, 2020; Edmonds, 2020; Edmonds and Candy, 2002).

Section 2 discusses the strategies for achieving Reliable, Safe & Trustworthy systems. Section 3 lays out the novel two-dimensional HCAI framework, describing different design objectives and the path to high levels of human control and high levels of automation. Section 4 provides design principles plus examples and Section 5 summarizes with limitations and conclusions.

2. **Reliable, Safe & Trustworthy Systems**

My discussion of Reliable, Safe & Trustworthy clarifies what leads to high performance. These definitions emphasize (1) technical practices that support reliability, (2) management strategies that create cultures of safety, and (3) independent oversight structures that support trust.

*Reliable* (Modarres et al., 2016) systems come from appropriate technical practices, which support human responsibility (Canadian Government, 2019), fairness (O'Neil, 2016), and explainability (Du, Liu & Hu, 2019), such as:

- **A**udit trails and analysis tools to review failures & near misses

- **B**enchmark tests, which are widely accepted for verification & validation
- **C**ontinuous review of data quality & bias testing to cope with shifting contexts of use
- **D**esign strategies to build confidence across stakeholder communities

Reliability is advanced by studying past performances by way of detailed audit trails, often called flight data recorders, which have been so effective in civil aviation. Ample testing and analyses of training data promote reliable performance, as do many other common software engineering practices.

Cultures of *safety* (Guldenmund, 2000; Berry et al., 2016) are cultivated by open management strategies such as

- leadership commitment to safety
- invitations to report problems within the organization
- reviews of failures and near misses by internal review boards
- public reports of failures to promote discussion
- internal oversight boards for future plans and past practices

all of which guide continuous refinement of management strategies, training, operational practices, and root-cause failure analyses. Capability Maturity Models guide organizational design to support safety, but further discussion of these strategies is beyond this paper.

There is a long history of discussions to define *trust* such as Fukuyama (1995), who focused on social trust "within a community of regular, honest, and cooperative behavior, based on commonly shared norms, on the part of the members of that community." His community-based definition leads to raising trust by relying on respected independent oversight structures (Shneiderman, 2000, 2016) including:

- companies with proven capacity to build, operate, and maintain technology,
- professional organizations that develop effective voluntary guidelines and standards
    (International Standards Organization, IEEE, Robotics Industry Association)
- government agencies that regulate in ways that promote innovation
    (US Food & Drug Administration, Federal Aviation Administration, National Highway Traffic Safety Administration)
- non-governmental organizations that certify companies and products
    (Underwriters Laboratories, Better Business Bureau, Consumer's Union)
- accounting firms with demonstrated value in auditing companies
    (KPMG, Ernst & Young, Deloitte, PwC), and
- insurance companies that promote trust by compensating for failures.

These independent oversight structures, which need much deeper discussions, develop trust in products and services by providing independent review over design, operation, and maintenance. Since some systems that are trusted may not be *trustworthy*, I will use the stronger term.

There are many components to the supporting structures of RST systems, but the intuitive meanings of these terms should also be clear. Users of mature technologies such as elevators, cameras, home appliances, powered wheelchairs, or medical devices know when these devices are trusted, reliable, and safe. They appreciate the high levels of automation, but think of themselves as operating these devices in ways that give them control so as to accomplish their goals. Designers who adopt the HCAI mindset will emphasize strategies for enabling diverse users to steer, operate, and control their highly automated devices, while inviting users to exercise their creativity to refine designs. Well-designed automation can ensure finer human finer control, such as in surgical robots that enable surgeons to make more precise incisions in difficult to reach organs.

Successful designs enable humans to work in interdisciplinary teams, so as to coordinate and collaborate with managers, peers, and subordinates. Computers are not teammates, collaborators, or co-active partners, as Johnson et al. (2014) suggest; computers should support human activities in ways that reduce workload, raise performance, and ensure safety. The larger design issues for RST systems also promote resilience, clarify responsibility, increase quality, and encourage creativity (Edmonds & Candy, 2002; Giuliani et al., 2010; Woods, 2017). Still broader goals are to ensure privacy, increase cybersecurity, support social justice, and protect the environment.

The HCAI framework presented in Section 3 guides designers and researchers to consider new possibilities that promote Reliable, Safe & Trustworthy systems. After all, most consumers, industrial supervisors, physicians, and airplane pilots are not interested in computer autonomy; what they want are RST systems to dramatically increase human performance, while simplifying their effort, so they can devote themselves to their higher aspirations. This paper shows how careful design leads to computer applications that promote human control while also providing high levels of automation.

3. The Human-Centered Artificial Intelligence for RST Systems

The HCAI framework in this paper steers designers and researchers to ask fresh questions and rethink the nature of automation/autonomy. As designers get beyond thinking of computers as our teammates, collaborators, or partners, they are more likely to develop technologies that dramatically increase human performance by taking advantage of unique computer features including sophisticated algorithms, advanced sensors, information abundant displays, and powerful effectors. Clarifying human responsibility also guides designers to support human capacity to invent creative solutions in novel contexts with incomplete knowledge. The HCAI framework clarifies that the goal of design excellence is to promote human self-efficacy, mastery, and responsibility, which support the Reliable, Safe & Trustworthy systems that managers and users seek.

The goals of RST are most relevant for life-critical systems, but there are many consumer and professional applications:

**Recommender systems:** these are widely used in advertising services, social media platforms, and search engines have brought strong benefits to consumers (Konstan & Riedl, 2012). Consequences of mistakes by these systems are usually less serious, possibly even giving consumers interesting suggestions of movies, books, or restaurants. However, malicious actors can manipulate these systems to influence buying habits, change election outcomes, spread hateful messages, and reshape attitudes about climate, vaccinations, gun control, etc. Thoughtful design that improves user control could increase consumer satisfaction and limit malicious use.

Other applications of automation include common user tasks, such as search query completion or spell checking. These tasks are carefully designed to preserve user control and avoid annoying disruptions, while offering useful assistance. Heer (2018) breaks new ground by offering ways of using automation in support of human control with examples from data cleaning, exploratory data visualization, and natural language translation.

**Consequential applications:** Moving up to the middle-range of consequential applications leads to medical, legal, environmental, or financial systems that can bring substantial benefits and harms. A well-documented case is the flawed Google Flu Trends, which was designed to predict flu outbreaks, enabling public health officials to assign resources more effectively (Lazer et al., 2013). The initial success did not persist and after two years, Google withdrew the website, because the programmers did not anticipate the many changes in search algorithms, user behavior, and societal context. Lazer et al. describe the harmful attitude of programmers as "algorithmic hubris," suggesting that some programmers have unreasonable expectations of their capacity to create foolproof autonomous systems, akin to what happened with the Boeing 737 MAX.

The financial flash crashes in public stock markets and currency exchanges caused by high frequency trading algorithms triggered losses of billions of dollars in just a few minutes. However, with adequate logs of trades, market managers can often repair the damage. Another saving grace for the middle-range of consequential applications in medical care, portfolio management, or legal advice, is that there may be time for alert decision makers to reflect on the algorithm's recommendation, consult with colleagues, or probe more deeply to understand the recommendation.

**Life-critical systems:** Moving up to the challenges of life-critical applications, we find the physical devices like self-driving cars, pacemakers, and implantable defibrillators, as well as complex systems such as military, industrial, and aviation applications. These applications require rapid actions and may have irreversible consequences.

The one-dimensional levels of automation, grew from Sheridan and Verplank's early paper, but was repeated in many articles and textbooks, including my own (Figure 1).

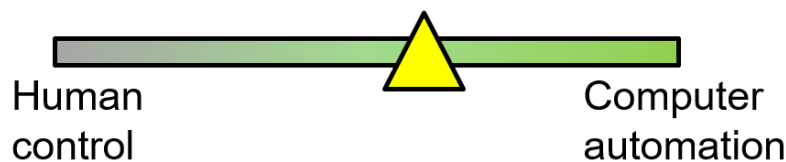

Figure 1: One-dimensional thinking suggest that designers must choose between human control and computer automation

Designers thought they had to choose a point on the one-dimensional line from human control to computer automation. The implicit message was that more automation meant less user control. The decoupling of these concepts leads to a two-dimensional HCAI framework, which suggests that achieving high levels of human control and high levels of automation is possible (yellow triangle in Figure 2).

The desired goal is often, but not always, to be in the upper right quadrant. Most RST systems are on the right side. The lower right quadrant is home to relatively well-understood and predictable tasks, e.g. automobile automatic transmission or skid control on normal highways. For poorly understood and complex tasks with varying contexts of use, the upper right quadrant is needed. These tasks involve creative decisions, making them currently at the research frontier. As contexts of use are standardized (e.g. elevator shafts) these tasks can become under greater computer control with high levels of automation.

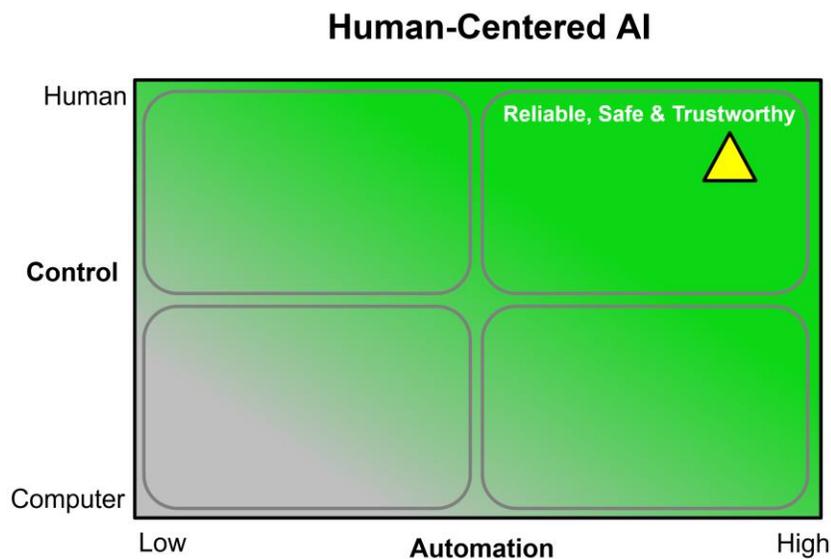

Figure 2: Two-dimensional framework with the goal of Reliable, Safe & Trustworthy, which is achieved by a high level of human control and high level of automation (yellow triangle).

The lower right quadrant (Figure 3), with high automation and low human control, is home of computer autonomy requiring *rapid action*, for example, airbag deployment, anti-lock brakes, pacemakers, implantable defibrillators, or defensive weapons systems. In these applications, there is no time for human intervention or control. Because the price of failure is so high, these applications require extremely careful design, extensive testing, and monitoring during usage at scale to refine designs. Effective and proven designs can become RST systems with higher levels of automation and less human supervision.

The upper left quadrant, with high human control and low automation, is the home of human autonomy where *human* mastery is desired to enable competence building, free exploration, and creativity. Examples include bicycle riding, piano playing, baking, or playing with children.

During these activities, humans generally want to derive pleasure from seeking mastery, improving their skills, and feeling fully engaged. They may elect to use computer-based systems for training, review, or guidance, but many people desire independent action to achieve mastery that builds self-efficacy. In these actions, the goal is in the doing and the personal satisfaction that it provides, plus the potential for creative exploration (Calvo et al., 2020).

The lower left quadrant is the home of simple devices such as clocks or mousetraps, as well as deadly devices such as land mines.

The take away message for designers is that, for certain tasks, there is value in full computer control or full human mastery. However, the challenge is to develop effective and proven designs, supported by trusted social structure, reliable practices, and cultures of safety so as to acquire RST recognition.

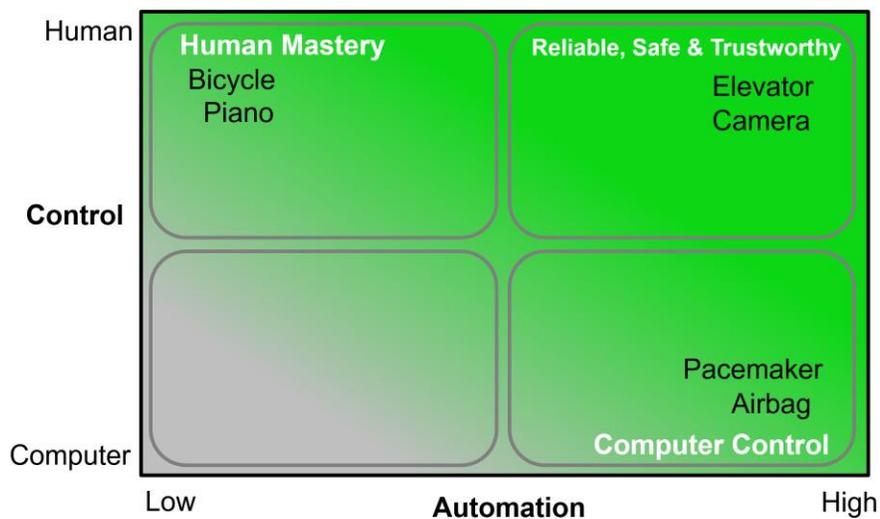

Figure 3: Regions requiring rapid action (high automation, low human control) and human mastery (high human control, low automation).

In addition to the edge cases of full computer and human autonomy, there are two other edge cases that signal danger -- *excessive automation* and *excessive human control*. On the far right (Figure 4) is the region of excessive automation, where there are dangers from designs such as the Boeing 737 MAX's MCAS system. Over and above the issues of single point of failure from one angle of attack sensor, the designers believed that their autonomous system could not fail. Therefore, its existence was not described in the user manual and the pilots were not trained in how to switch to manual override. The IBM AI Guidelines wisely warns that "imperceptible AI is not ethical AI" (https://www.ibm.com/design/ai/fundamentals/).

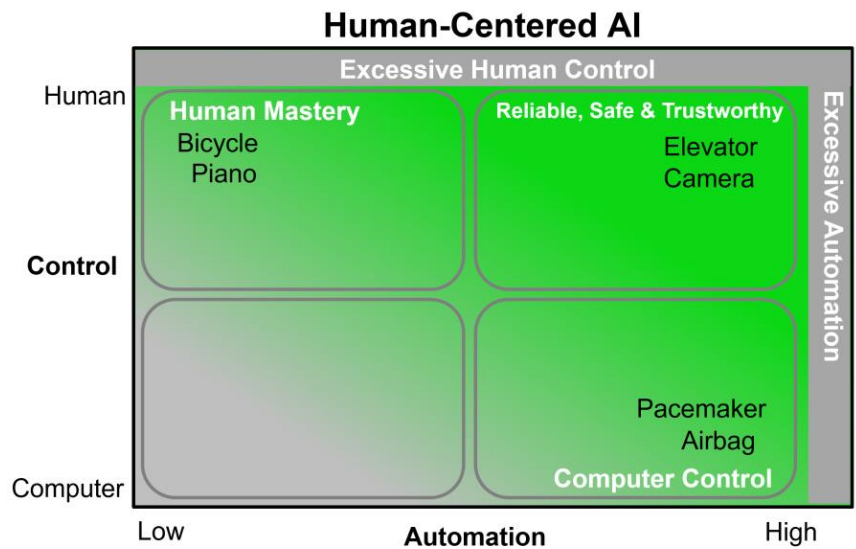

Figure 4: Designers need to prevent the failures from excessive automation and excessive human control (gray areas)

Excessive automation was cited by the National Transportation Safety Board's report on the deadly 2016 crash of a Tesla car (2017). That report cautioned that "...automation 'because we can' does not necessarily make the human-automation system work better. ... This crash is an example of what can happen when automation is introduced 'because we can' without adequate consideration of the human element." The Tesla "Autopilot" system name suggests greater capability than is available. This can encourage drivers to become less vigilant, because designers have allowed drivers to ignore safety warnings about paying attention to road conditions and other vehicles.

Similarly, excessive human control allows people to make deadly mistakes. Automation controls, human factors affordances, and prevention of catastrophic failures are established strategies for reducing the number of deadly outcomes from "human mistakes", which should be seen as design failures (Thimbleby, 2020). Self-driving car advocates stress the goal of preventing human errors that lead to accidents, but even modest changes to existing car designs can promote safety, such as collision avoidance to slow cars as they approach vehicles just in front of their current position. These modest changes can then be refined and adapted in ways that increase safety, eventually leading to less need for driver control, with greater supervisory control from regional centers, akin to air-traffic control centers, that manage traffic flow of thousands of cars, track frequent near misses, and make changes to road speeds depending on weather and other conditions.

Further ways of preventing excessive human control include car-based ignition interlock devices that conduct breathalyzer tests to prevent drivers with high levels of alcohol in their breath from turning on their cars. Positive Train Control systems limit engineers from going at high speeds on curves or in terminal areas. Numerous interlocks in aviation prevent pilots from making mistakes, for example, reverse thrusters on jet engines can only be engaged once the strain gauge in the landing gear indicates that the plane is on the ground.

Home appliances also have guards, such as the interlock on self-cleaning ovens to prevent homeowners from opening oven doors when temperatures are above 600F degrees. Poorly designed interlocks can be annoying, such as the overly ambitious automobile systems that lock doors too often. Software-based constraints, such as range checking, generalize the interlocks and guards, ensuring that algorithms have permissible inputs and producing only acceptable outcomes.

Interlocks or guards are helpful in preventing mistakes, but additional design and monitoring features are needs to ensure TRS designs. The challenging question is how to integrate HCAI designs into life-critical systems like aviation, flexible manufacturing systems, implantable pacemakers, and automobiles. There is huge current interest in self-driving, driverless, or autonomous cars, which are sometimes described as allowing passengers to sit back to read or sleep, while the car takes them safely to their destination. This seductive vision is widely appealing, especially for its potential improved safety, but there is a growing realization that this may be an overstatement of what is possible in the next decade. Most automobile manufacturers are wisely proceeding cautiously. They started with anti-lock brakes and cruise control decades ago, and now offer features like lane control, parking assist, auto-braking, and collision avoidance. This incremental approach supports development and refinement processes that eventually lead to RST designs that gain widespread acceptane.

The notion of full car autonomy is productively giving way to a human-centered approach, which provides appropriate controls to improve safety, while addressing the numerous special cases of snow-covered roads, unusual vehicles, vandalism, malicious interference, construction zones, and emergency guidance from police, fire, or ambulance personnel. As car sensors, highway infrastructure, and car-to-car communication improve, strategies to make cars as safe as elevators will become easier to implement.

Figure 5 shows the relative positions of 1980 cars, 2020 cars, self-driving cars, and the proposed goal of Reliable, Safe & Trustworthy(RST) cars in 2040.

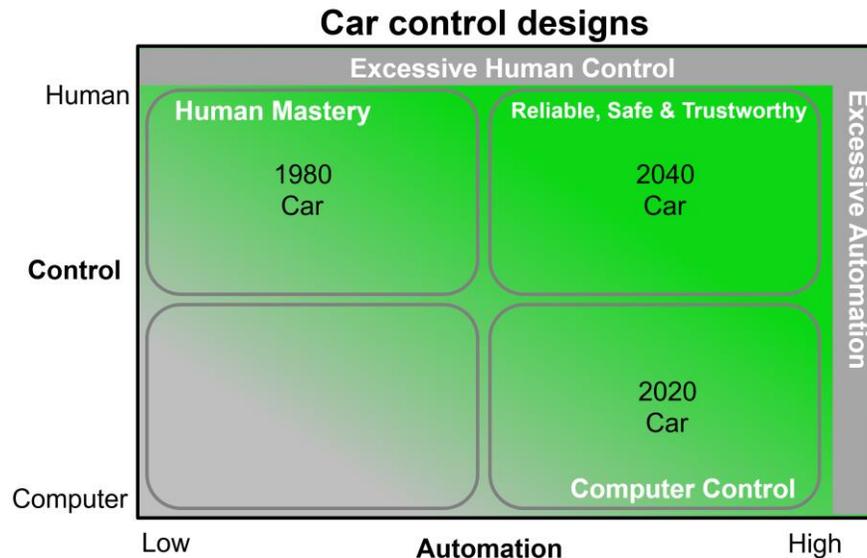

Figure 5: 1980 cars had modest automation with high levels of human control, while 2020 self-driving cars have high automation, but inadequate human control.  Achieving Reliable, Safe & Trustworthy (RST) self-driving cars by 2040 could be done with high automation and high human control, while avoiding the dangers of excessive automation or excessive human control

The four quadrants may be helpful in suggesting differing designs for a product or service:

**Example 1:** Patient Controlled Analgesia (PCA) devices (Figure 6) allow post-surgical, severe cancer, or hospice patients to select the amount and frequency of pain control medication. There are dangers and problems with young and old patients, but with good design and management PCA devices deliver safe and effective pain control (Macintyre, 2001). A simple morphine drip bag design for the lower left quadrant (low automation and low human control) would regularly deliver a fixed amount of pain control medication. A more automated design for the lower right quadrant (increased automation, but little human control) would provide machine selected doses that could vary by time of day, activity, and data from body sign sensors, although these do not assess perceived pain.

A human-centered design for the upper left quadrant (higher human control, with low automation) would allow patients to squeeze a trigger to control the dosing, frequency, and total amount of pain control medication. However, the dangers of overdosing must be controlled by an interlock that prevents frequent doses, typically with lockout periods of 6-10 minutes, and total dose limits over 1-4 hour periods. Finally, a RST design for the upper right quadrant would allow users to squeeze a trigger to get more pain medicine, but would use sensors, and machine learning to choose appropriate doses based on patient and disease variables, while preventing overdosing. Patients would be able to get information on why limiting pain medication is important with explanations of how they operate the PCA device. The RST design would include a hospital control center to monitor usage of hundreds of PCA devices so as to ensure safe practices, deal with power or other failures, review audit trails, and collect data to improve the next generation of PCA devices.

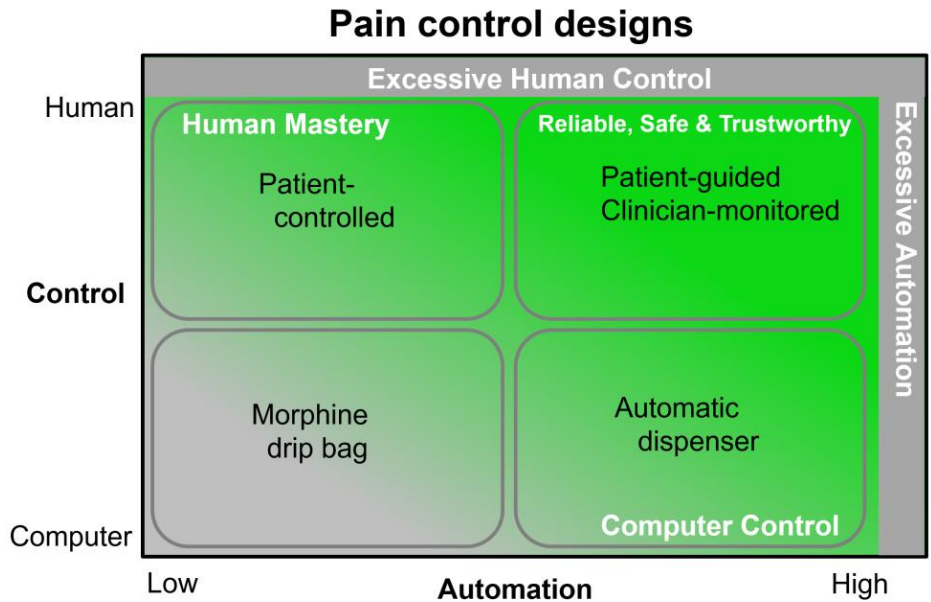
Figure 6: Four approaches to pain control designs

4. **Prometheus Principles and Examples**

This new goal of HCAI overcomes the 50-year debate, lucidly described by Markoff (2016), between those who sought artificial intelligence and those who sought intelligence augmentation (IA). This debate over AI vs IA now seems like arguing about iron horses vs horses or how many angels can fit on a pinhead. In fact, IBM recently declared "There's no AI without IA". Designers can produce HCAI by integrating artificial intelligence algorithms with user interface designs in ways that amplify, augment, enhance, and empower people.

The current version of the Google design guidebook (https://pair.withgoogle.com/intro/) buys into to the need to choose: "Assess automation vs. augmentation... One large consideration is if you should use AI to automate a task or to augment a person's ability to do that task themselves... For AI-driven products, there's an essential balance between automation and user control." However, the authors of the Google design guidebook are open to the possibility that "When done right, automation and augmentation work together to both simplify and improve the outcome of a long, complicated process."

Other guidelines (Horvitz, 1999; Dudley & Kristensson, 2018) paved the way for Microsoft's Guidelines for AI-Human Interaction (https://www.microsoft.com/en-us/research/project/guidelines-for-human-ai-interaction/), which has 18 guidelines for initial use, normal use, coping with problems, and changes over time (Amershi et al., 2019). These guidelines emphasize user understanding and control, while addressing ways for the system to "make clear why the system did what it did" and "learn from user behavior."

The IBM Design for AI website (https://www.ibm.com/design/ai/fundamentals/) discusses issues more broadly, offering a high-level tutorial that covers design foundations, technology basics, ethical concerns, accountability, explainability, and fairness, but I question its suggestion that a system should "endear itself to the user" and "form a full emotional bond."

Endsley's "Guidelines for the design of human-autonomy systems" (2018) has 20 thoughtful items that cover human understanding of autonomous systems, minimizing complexity, and supporting situation awareness. Typical guidelines are: "Use automated assistance for carrying out routine tasks rather than higher-level cognitive functions" and "Provide automation transparency," each of which is explained in detail.

Successful designs are comprehensible, predictable, and controllable, thereby increasing the users' self-efficacy, leading to Reliable, Safe & Trustworthy systems. These successes require careful design of the fine structure of interaction, which emerges from validated theories, clear principles, and actionable guidelines. In turn, this knowledge is embedded in programming tools that support human control over complex processes. I'll call these design rules the Prometheus Principles, after the Greek god who gave fire as a gift to humanity. These potent principles include:

- consistent interfaces to allow users to form, express, and revise intent,
- continuous visual display of the objects and actions of interest,
- rapid, incremental, and reversible actions,
- informative feedback to acknowledge each user action,
- progress indicators to show status, and
- completion reports to confirm accomplishment.

The design decisions to craft user interfaces based on the Prometheus Principles typically involve tradeoffs (Fischer, 2018), so careful study, creative design, and rigorous testing are needed. Then implemented designs must be continuously monitored by domain experts to understand problems and refine the designs.

The examples that follow clarify the key ideas of supporting users to express their intent through visual interfaces (although auditory and haptic interfaces are also valuable in some situations and for users with disabilities) and to get informative feedback about the machine state, with progress indicators and completion reports to let users know status. Many real-world designs are flawed, but these positive examples describe how it is possible to support both high levels of human control and high levels of automation:

**Example 2:** Simple thermostats allow residents to take better control of the temperature in their homes. They can see the room temperature and the current thermostat setting, clarifying what they should do to get a higher/lower setting. Then they can hear the heating system turn on/off or see a light to indicate that their action has produced a response. The basic idea is to give residents an awareness of the current state, allow them to reset the control, and then give informative feedback that the computer is acting on their intent. There may be further feedback as the residents see the thermometer rise in response to their action, and maybe an indication when their desired goal has been achieved. Thermostats offer still further benefits – they continue to keep the room temperature at the new setting automatically. In summary, while some thermostats may lack the features necessary that give clear feedback, well-designed thermostats give users an understanding of how they have controlled the automation to get the temperature they desire in their homes. Newer programmable thermostats apply machine learning to allow residents to better accommodate their schedules and save energy.

**Example 3:** Small automations are part of many successful products, such as automobiles. While power steering and automatic transmissions are mature automations that are widely accepted, anti-lock brake systems apply the same principles to enable drivers to maintain traction during challenging road conditions. Cruise control and adaptive cruise control are more complex, but they give drivers better control over the speed of their vehicles. Newer automations that augment human abilities to increase safety include parking assistance, lane following, and collision avoidance. Current technology, such as Tesla's Autopilot, are "designed to be able to conduct short and long distance trips with no action required by the person in the driver's seat… in almost all situations", but regulatory approval, independent oversight mechanisms, and demonstration of safety are still needed to achieve RST status. Tesla cautions that "current Autopilot features require active driver supervision and do not make the vehicle autonomous" (https://www.tesla.com/autopilot). The promise of advanced automation is to save lives, but critics have suggested that Tesla's unfortunate choice of the term Autopilot may encourage drivers to lower their vigilance, thereby leading to more accidents.

**Example 4:** Home appliances, such as dishwashers, clothes washers/dryers, and ovens allow users to choose settings that specify what they want, and then turn control over to sensors and timers to govern the process. When well-designed, these machines offer users comprehensible and predictable controls that allow them to express their intent with controls that let them stop dishwashers to put in another plate or change from baking to broiling to brown their chicken. These automations give users better control of these appliances to ensure that they get what they want.

**Example 5:** Well-designed user interfaces in elevators enable substantial automation while providing appropriate human control. Elevator users approach a simple two-button control panel and press up or down. The button lights to indicate that the users' intent has been recognized. A display panel indicates the elevator's current floor so users can see progress, which lets them know how long they will have to wait. The elevator doors open, a tone sounds, and the users can step in to press a button to indicate their choice of floors. The button lights up to indicate their intent is recognized and the door closes. The floor display shows progress towards the goal. On arrival, a tone sounds and the door opens. The automated design which replaced the human operator ensures that doors will only open while on a floor, while triply redundant safety systems prevent elevators from falling, even under loss of power or broken cables. Automation algorithms coordinate multiple elevators, automatically adjusting their placement based on time of day and changing passenger loads. Override controls allow firefighters or moving crews to achieve their goals. There are many refinements, such as detectors to prevent doors from closing on passengers, so that the overall design is Reliable, Safe & Trustworthy.

**Example 6:** Digital cameras in most cell phones display an image of what the users would get if they clicked on the large button. The image is updated smoothly as users adjust their position or zoom in. At the same time, the camera makes automatic adjustments to the aperture and focus, while compensating for shaking hands, a wide range of lighting conditions (high dynamic range), and many other factors. Flash can be set to be on or off, or left to the camera to choose automatically. Users can choose portrait modes, panorama, and video, including slow motion and time lapse. Users also can set various filters and once the image is taken they can make further adjustments such as brightness, contrast, cropping, and red-eye elimination. These designs give users high degree of control while also providing a high level of automation. Of course, there are mistakes, such as when the automatic focus feature puts the focus on a nearby bush, rather than the person standing just behind it. However, knowledgeable users can touch the desired focus point to avoid this mistake.

The Prometheus Principles help designers develop the comprehensible, predictable, and controllable interfaces that support RST systems. Many of the principles are embedded in Human Interface Guidelines documents, such as Apple's, which stipulates that "User Control: … people - not apps - are in control" and "Flexibility: … (give) users complete, fine-grained control over their work." (Apple 2019). These guidelines and others (Shneiderman et al., 2016) that emphasize human control offer very different advice than the simplistic pursuit of computer autonomy. Designs based on these Prometheus Principles can lead to, what I believe is desired by most users: Reliable, Safe & Trustworthy systems.

## 5. Summary, limitations, and conclusions

The HCAI framework separates the issue of human control from computer automation, making it clear that high levels of human control and high levels of automation can be achieved by good design. The design decisions give human operators a clear understanding of the machine state and their choices, guided by concerns such as the consequences and reversibility of mistakes. Well-designed automation, preserves human control where appropriate, thereby increasing performance and enabling creative improvements.

The HCAI framework suggests when computer control for rapid automated action is necessary, when human desire for mastery is paramount, and when there are dangers of excessive automation or excessive human control. It clarifies design choices for (1) consumer and professional applications, such as widely-used recommender systems, advertising tools, social media platforms, and search engines, which have brought strong benefits to consumers, (2) consequential applications in medical, legal, environmental, or financial systems that can bring substantial benefits and harms, and (3) life-critical applications such as cars, airplanes, trains, military systems, pacemakers, or intensive care units.

The HCAI framework is based on the belief that people are different from computers. Therefore, designs which take advantage of unique computer features including sophisticated algorithms, advanced sensors, information abundant displays, and powerful effectors are more likely to increase performance. Similarly designs and organizational structures which recognize the unique capabilities of humans will have advantages such as encouraging innovative use, supporting continuous improvement, and promoting breakthroughs to vastly improved designs.

An important research direction is to develop objective measures of the levels of control and autonomy, tied to diverse tasks. Such measures would stimulate more meaningful design discussions, which would lead to improved guidelines, evaluations, and theories.

Human responsibility for mistakes is another powerful driver of design advancements such as the inclusion of detailed audit trails and consistent informative feedback about machine state. Then retraining of users and redesign of systems happens rapidly. Difficult questions remain, such as dealing with the de-skilling effects that undermine human skills that may be needed when automation fails, and the difficulty of remaining vigilant when user actions become less frequent.

Ethical questions, such as considerations of responsibility, fairness, and explainability, are helpful in developing general principles. When these general principles are combined with deep knowledge and experience with the complexities of product and service design, they can yield actionable guidelines (IEEE, 2019). The HCAI framework lays a foundation for responsibility, fairness, and explainability, but actionable guidelines to achieve these goals are beyond this paper.

A human-centered RST system, will be (1) reliable because of sound technical practices, (2) safe because of open management strategies that build cultures of safety, and (3) trustworthy because of well-designed independent oversight structures. RST systems will evolve rapidly because mechanisms to monitor failures and near misses support creative improvements.

The goal of this paper is to encourage artificial intelligence researchers and developers who design products and services to shift from one-dimensional thinking about levels of automation/autonomy to a fresh two-dimensional HCAI framework. The HCAI framework guides researchers and designers in designing technologies to give users appropriate control, while providing high levels of automation. When successful these technologies amplify, augment, enhance, and empower users, giving them dramatic increases in their performance.

**Acknowledgements**: Thanks to Michael Bernstein, Linda Candy, Ryan Carrier, Ernest Edmonds, Robert Fraser, Harry Hochheiser, Robert Hoffman, Eric Hughes, Gary Klein, Alan Mackworth, Doug Oard, Catherine Plaisant, Jennifer Preece, Robin Murphy, Steven M. Rosen, Arnon Rosenthal, Ben Sawyer, Thomas Sheridan, Mark Smith, Harold Thimbleby, Fernanda Viegas, Jamie Waese, Martin Wattenberg, and David D. Woods for comments on early drafts.

**References**

Amershi, S., Weld, D., Vorvoreanu, M., Fourney, A., Nushi, B., Collisson, P., ... & Horvitz, E. (2019). Guidelines for human-AI interaction. *Proceedings of the 2019 CHI Conference on Human Factors in Computing Systems* (Paper 3, 1-13). ACM.

Apple Computer, Inc., Human Interface Guidelines, Apple, Cupertino, CA (2019). Available at https://developer.apple.com/design/human-interface-guidelines/ios/overview/themes/

Bainbridge, L. (1983). Ironies of automation, *Automatica,* 19, 6, 775-779, Pergamon.

Bennett, K. B. & Hoffman, R. R. (2015). Principles for interaction design, Part 3: Spanning the creativity gap. *IEEE Intelligent Systems, 30* (6), 82-91.

Berry, J. C., Davis, J. T., Bartman, T., Hafer, C. C., Lieb, L. M., Khan, N. & Brilli, R. J. (2016). Improved safety culture and teamwork climate are associated with decreases in patient harm and hospital mortality across a hospital system. *Journal of Patient Safety* (Jan 7 2016). http://www.ncbi.nlm.nih.gov/pubmed/26741790

Blackhurst, J. L., Gresham, J. S., & Stone, M. O. (2011). The autonomy paradox. *Armed Forces Journal*, 20-40.

Bradshaw, Jeffrey M., Robert R. Hoffman, David D. Woods & Matthew Johnson (2013). The seven deadly myths of autonomous systems. *IEEE Intelligent Systems* 28, 3, 54-61.

Brooks, R. (2017). The big problem with self-driving cars is people. *IEEE Spectrum: Technology, Engineering, and Science News* (27 July 2017).

Calvo RA, Peters D., Vold V, Ryan, RM (2020). Supporting human autonomy in AI systems: A framework for ethical enquiry. In Burr, C. & Floridi, L. (Eds.) *Ethics of Digital Well-Being: A Multidisciplinary Approach*. Springer Open.

Canadian Government (2019). Responsible use of artificial intelligence (AI), https://www.canada.ca/en/government/system/digital-government/modern-emerging-technologies/responsible-use-ai.html

Candy, L. (2020). Creating with the Digital: Tool, Medium, Mediator, Partner, In Brooks, A. L. (editor) *Interactivity, Game Creation, Design, Learning, and Innovation*. Springer (to appear). http://lindacandy.com/wp-content/uploads/2019/12/ArtsIT-LCandy.pdf

de Visser, E. J., Pak, R. & Shaw, T. H. (2018). From 'automation' to 'autonomy': the importance of trust repair in human–machine interaction. *Ergonomics, 61*(10), 1409-1427.

Defense Science Board (2016). "Summer Study on Autonomy." Office of the Undersecretary for Defense for Acquisition, Technology and Logistics, Department of Defense, Washington, DC.

Du, M., Liu, N., & Hu, X. (2020). Techniques for interpretable machine learning. *Communications of the ACM, 63* (1), 68-77.

Dudley, J. J., & Kristensson, P. O. (2018). A review of user interface design for interactive machine learning. *ACM Transactions on Interactive Intelligent Systems (TiiS),* 8(2), 8.

Edmonds, E. A. (2020). Computation for Creativity. In J. S. Gero and M. L Maher (Editors), *Computational and Cognitive Models of Creative Design*, Springer, London, (to appear).

Edmonds, E. A. and Candy, L. (2002). Creativity, Art Practice and Knowledge, *Communications of the ACM* Special Section on Creativity and Interface, 45(10), 91-95.

Endsley, M. R. (2017). From here to autonomy: lessons learned from human–automation research, *Human Factors, 59* (1), 5-27.

Endsley, Mica R. (2018). Level of Automation Forms a Key Aspect of Autonomy Design, *Journal of Cognitive Engineering and Decision Making* 12, 1 (March 2018), 29-34, DOI: 10.1177/1555343417723432

Fischer, G. (2018). Design Trade-Offs for Quality of Life. *ACM Interactions XXV*, 1, 26-33.

Fukuyama, F. (1995). *Trust: The Social Virtues and the Creation of Prosperity*. Free Press, New York.

Giuliani, M., Lenz, C., Müller, T., Rickert, M., & Knoll, A. (2010). Design principles for safety in human-robot interaction. *International Journal of Social Robotics, 2* (3), 253-274.

Guldenmund, F. W. (2000). The nature of safety culture: a review of theory and research. *Safety Science, 34* (1-3), 215-257.

Hancock, P. A. (2017). Imposing limits on autonomous systems. *Ergonomics, 60* (2), 284-291.

Heer, Jeffrey (2019) Agency plus automation: Designing artificial intelligence into interactive systems, *Proc. National Academy of Sciences*, https://www.pnas.org/content/116/6/1844

Hoffman, R. R., Cullen, T. M. & Hawley, J. K. (2016). The myths and costs of autonomous weapon systems. *Bulletin of the Atomic Scientists, 72* (4), 247-255.

Hoffman, R. R. & Johnson, M. (2019). The quest for alternatives to "Levels of Automation" and "Task Allocation." In M. Mouloua & P.A. Hancock (Eds.), *Human Performance in Automated and Autonomous Systems*. Boca Raton, FL: CRC Press, 43-68.

Horvitz, E. (1999). Principles of mixed-initiative user interfaces. *Proceedings of the SIGCHI Conference on Human Factors in Computing Systems*. ACM, 159-166.

IEEE Global Initiative on Ethics of Autonomous and Intelligent Systems (2019). *Ethically Aligned Design: A Vision for Prioritizing Human Well-being with Autonomous and Intelligent Systems, First Edition.* IEEE. https://standards.ieee.org/content/ieee-standards/en/industry-connections/ec/autonomous-systems.html

Johnson, M., Bradshaw, J. M., & Feltovich, P. J. (2018). Tomorrow's human–machine design tools: From levels of automation to interdependencies. *Journal of Cognitive Engineering and Decision Making, 12*(1), 77-82.

Johnson, M., Bradshaw, J. M., Feltovich, P. J., Jonker, C. M., Van Riemsdijk, M. B., & Sierhuis, M. (2014). Coactive design: Designing support for interdependence in joint activity. *Journal of Human-Robot Interaction*, 3(1), 43-69.

Jordan M. I. (2018). Artificial intelligence - The revolution hasn't happened yet. https://medium.com/@mijordan3/artificial-intelligence-the-revolution-hasnt-happened-yet-5e1d5812e1e7

Kaber, D. B. (2018). Issues in human–automation interaction modeling: Presumptive aspects of frameworks of types and levels of automation, *Journal of Cognitive Engineering and Decision Making,* 12 (1), 7-24.

Klein, G., Woods, D.D., Bradshaw, J.M., Hoffman, R.R. & Feltovich, P.J. (2004). Ten challenges for making automation a "team player" in joint human-agent activity. *IEEE Intelligent Systems* **6**, 91–95.

Konstan, J. A., & Riedl, J. (2012). Recommender systems: from algorithms tommen reco user experience. *User Modeling and User-Adapted Interaction, 22* (1-2), 101-123.

Lazer, D., Kennedy, R., King, G. & Vespignani, A. (2014). The parable of Google Flu: Traps in the big data analysis, *Science 343* (March 14, 2014), 1203-1205.

Li, F.-F. (2018). How to make A.I. that's good for people. *The New York Times* (March 7, 2018). https://www.nytimes.com/2018/03/07/opinion/artificial-intelligence-human.html

Macintyre, P. E. (2001). Safety and ef®cacy of patient-controlled analgesia, *British Journal of Anaesthesia 87*, 1, 36-46.

Markoff, J. (2016). *Machines of Loving Grace: The Quest for Common Ground between Humans and Robots*, HarperCollins Publishers.

Mindell, D. (2015). *Our Robots, Ourselves: Robotics and the Myths of Autonomy,* Viking Press.

Modarres, M., Kaminskiy, M. P., & Krivtsov, V. (2016). *Reliability Engineering and Risk Analysis: A Practical Guide*. CRC press.

Murphy, R. and Shields, J. (2012). The Role of Autonomy in DoD Systems, Defense Science Board Task Force Report (July 2012), Washington, DC.

Nicas, J., Kitroeff, N., Gelles, D., & Glanz, J. (2019). Boeing Built Deadly Assumptions into 737 Max, Blind to a Late Design Change. *The New York Times* (June 6, 2019). https://www.nytimes.com/2019/06/01/business/boeing-737-max-crash.html

O'Neil, C. (2016). *Weapons of Math Destruction: How Big Data Increases Inequality and Threatens Democracy*, Crown Publishers, New York.

Parasuraman, R., Sheridan, T. B., & Wickens, C. D. (2000). A model for types and levels of human interaction with automation. *IEEE Transactions on Systems, Man and Cybernetics-Part A: Systems and Humans, 30*: 286–297.

Sheridan, T. B. (1992). *Telerobotics, Automation, and Human Supervisory Control.* MIT Press.

Sheridan, T. B. (2000). Function allocation: algorithm, alchemy or apostasy? *International Journal of Human-Computer Studies, 52* (2), 203-216.

Sheridan, T. B., & Verplank, W. L. (1978). Human and computer control of undersea teleoperators. Massachusetts Institute of Technology Cambridge Man-Machine Systems Lab.

Shneiderman, B. (1983). Direct Manipulation: A Step Beyond Programming Languages. *IEEE Computer 8*, 57-69.

Shneiderman, B. (1987). *Designing the User Interface: Strategies for Effective Human-Computer Interaction*, Addison-Wesley Publ. Co., Reading, MA.

Shneiderman, B. (2000). Designing trust into online experiences. *Communications of the ACM, 43* (12), 57-59.

Shneiderman, B. (2016). Opinion: The dangers of faulty, biased, or malicious algorithms requires independent oversight. *Proceedings of the National Academy of Sciences, 113* (48), 13538-13540.

Shneiderman, B., Plaisant, C., Cohen, M., Jacobs, S. & Elmqvist, N. (May 2016). *Designing the User Interface: Strategies for Effective Human-Computer Interaction: Sixth Edition*, Pearson.

Society of Automotive Engineers (2014). Taxonomy and Definitions for Terms Related to On-Road Motor Vehicle Automated Driving Systems. SAE Report J3016. https://www.sae.org/standards/content/j3016_201401/

Strauch, B. (2017). Ironies of automation: Still unresolved after all these years. *IEEE Transactions on Human-Machine Systems, 48* (5), 419-433.

Thimbleby, H. (2020). *Fix IT: Stories from Healthcare IT*, Oxford University Press.

U.S. National Transportation Safety Board (2017). Collision Between a Car Operating With Automated Vehicle Control Systems and a Tractor-Semitrailer Truck Near Williston, Florida May 7, 2016, Report HAR1702.

Woods, D. D. (2017). Essential characteristics of resilience. In Hollnagel, E., Woods, D. W., and Leveson, N. (editors), *Resilience Engineering: Concepts and Precepts,* Ashgate Publishing, 21-34.

Woods, D. D., Tittle, J., Feil, M., & Roesler, A. (2004). Envisioning human-robot coordination in future operations. *IEEE Transactions on Systems, Man, and Cybernetics, Part C (Applications and Reviews), 34* (2), 210-218.